\documentclass[11pt,preprint]{aastex}
\usepackage{graphics,graphicx}
\usepackage{longtable}
\newcommand{\beq}{\begin{eqnarray}}
\newcommand{\eeq}{\end{eqnarray}}

\makeatletter
\renewcommand{\section}{\@startsection%
{section}{1}{0mm}{-\baselineskip}%
{0.5\baselineskip}{\normalfont\Large\bfseries}}%
\makeatother

\begin{document}


\pagestyle{plain}
\pagenumbering{arabic}

\large
\begin{center}
{\bf Optimizing Multi-Wavelength Blazar Studies through {\it Fermi}-LAT and {\it Swift} Synergy}\\
\normalsize
\vspace{+0.025in}
Christina D. Moraitis (Samford University, NASA Goddard Space Flight Center), D. J. Thompson (NASA Goddard Space Flight Center)
\end{center}

\vspace{+0.03in}
\normalsize





\vspace{-0.2cm}

\small{
\noindent 
} 
\normalsize
\normalsize


\smallskip
\noindent {\bf Abstract}
\vspace{+0.025in}

Blazar flares seen by the {\it Fermi Gamma-Ray Space Telescope} Large Area Telescope ({\it Fermi }LAT) are often followed up by Target of Opportunity (ToO) requests to the {\it Neil Gehrels Swift Observatory (Swift)}.  Using flares identified in the daily light curves of {\it Fermi } LAT Monitored Sources, we investigated which follow-up {\it Swift} ToO requests resulted in refereed publications.  The goal was to create  criteria of what {\it Swift} should look for in following up a {\it Fermi}-LAT gamma-ray flare.  Parameters tested were peak gamma-ray flux, flare duration (based on a Bayesian Block analysis), type of AGN (BL Lac or FSRQ), and pattern of activity (single flare or extensive activity). We found that historically active sources and high-photon-flux sources result in more publications, deeming these successful {\it Swift} ToOs, while flare duration and type of AGN had little or no impact on whether or not a ToO led to a publication.

\smallskip
\noindent {\bf 1. Introduction}
\vspace{+0.025in}

The {\it Fermi Gamma-Ray Space Telescope} \citep{atwood09} and the {\it Neil Gehrels Swift Observatory (Swift)} \citep{Gehrels2004} are both key tools for studying flaring gamma-ray sources. {\it Fermi} LAT has a broad observation range, from $<$0.1-$>$300 GeV energies. Not only so, but through LAT's unprecedented wide field of view, it has observed the entire celestial sphere regularly for over ten years of its mission. Often when a flare is detected with {\it Fermi} LAT, that source will be followed up via a Target of Opportunity (ToO) request to {\it Swift} to further study the source in X-ray, UV and optical wavelengths. Doing so will tell more about the source, what specific type of blazar it may be, and what physical processes are occurring to cause the observed flares. As more and more gamma-ray sources are observed by {\it Fermi}, more and more ToOs are sent  to {\it Swift} for multi-wavelength follow up. There will be a  time when {\it Swift} will not be able to follow up all of the ToO requests that it receives. Due to this potential saturation of requests, a study was conducted as a student project over the course of two months to help optimize future ToO requests to {\it Swift} with the goal of developing guidelines for when a ToO is most likely to be fruitful.

\begin{figure*}[ttt!]
\vspace{-0.3cm}
\hspace{-0.6cm}
\hbox{
\includegraphics[width=0.9\textwidth]{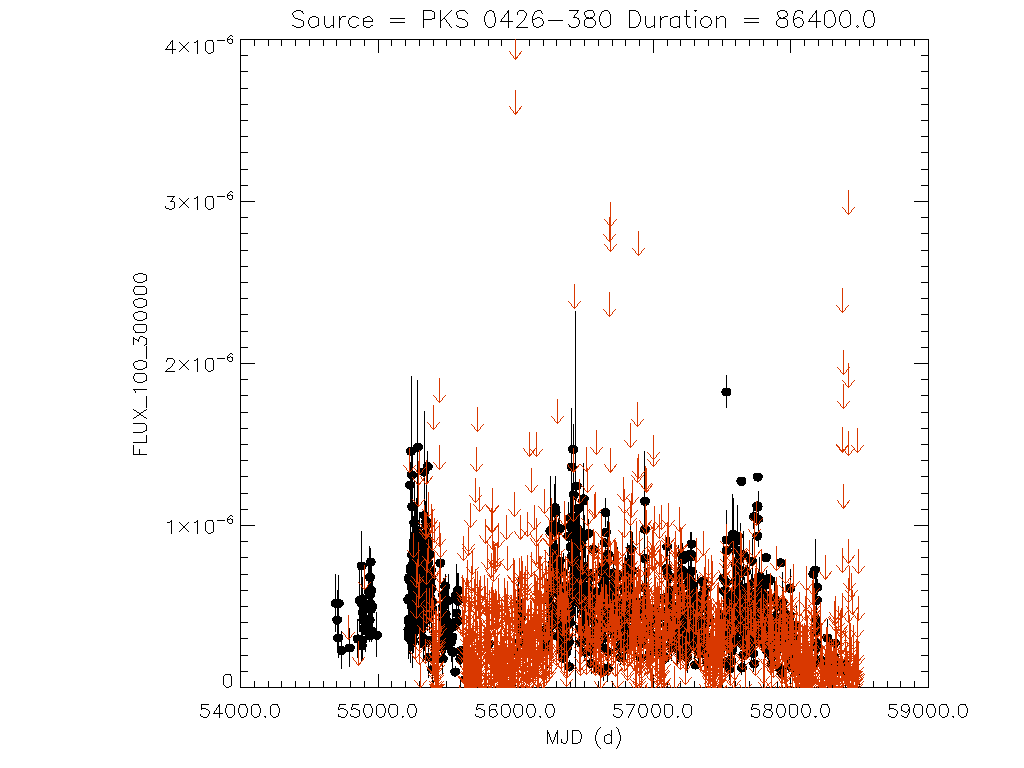}
}
\caption
{\it Example of a daily light curve for a gamma-ray blazar, from the  {\it Fermi}-LAT Monitored Source List.  Flux is given in units of $10^{-6} \rm{ph}~\rm{cm}^{-2}~s^{-1}$ above 100 MeV as a function of Modified Julian Day.  Arrows represent upper limits for days in which the source was not significantly detected. }
\label{fig:Monitored2}
\end{figure*}

\begin{figure*}[ttt!]
\vspace{-0.3cm}
\hspace{-0.4cm}
\hbox{
\includegraphics[width=0.9\textwidth]{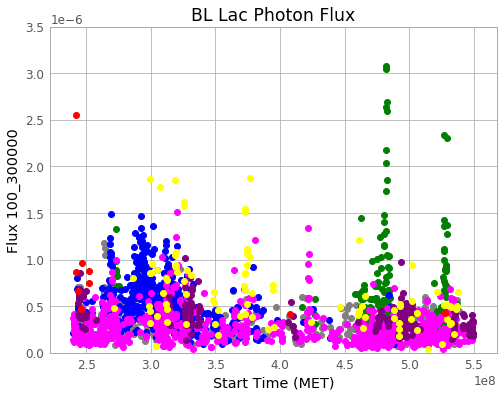}
}
\caption
{\it Example of light curves of several BL Lac objects from the  {\it Fermi}-LAT Monitored Source List.  Flux is given in units of $10^{-6} \rm{ph}~\rm{cm}^{-2}~s^{-1}$ above 100 MeV as a function of Mission Elapsed Time (MET), defined as seconds from 2001.0 UTC.}
\label{fig:BLLacs}
\end{figure*}

\smallskip
\noindent {\bf 2. Method}
\vspace{+0.025in}

The starting point for this project was the {\it Fermi}-LAT Monitored Source List, available from the  {\it Fermi} Science support Center Web site: {\it https://fermi.gsfc.nasa.gov/ssc/data/access/lat/msl\_lc/}.  This public data set provides daily and weekly light curves generated by an automated analysis, for all gamma-ray sources whose daily flux has ever exceeded $1 \times 10^{-6} \rm{ph}~\rm{cm}^{-2}~s^{-1}$ above 100 MeV.  Although these light curves do not have absolute flux calibration consistent over the life of the mission, they offer a convenient way to identify short-term flaring activity.  The vast majority of the sources in this list are gamma-ray blazars.  An example is shown in Fig. \ref{fig:Monitored2}.

\begin{figure*}[ttt!]
\vspace{-0.3cm}
\hspace{-0.6cm}
\hbox{
\includegraphics[width=0.9\textwidth]{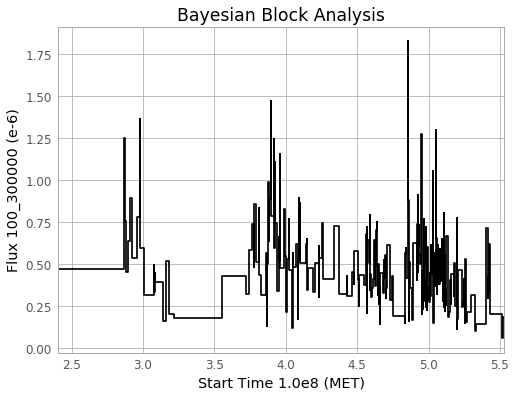}
}
\caption
{\it Bayesian Block analysis of a daily light curve for a gamma-ray blazar, from the  {\it Fermi}-LAT Monitored Source List.  Flux is given in units of $10^{-6} \rm{ph}~\rm{cm}^{-2}~s^{-1}$ above 100 MeV as a function of MET.   }
\label{fig:Bayesian}
\end{figure*}

The analysis steps are as follows:

\begin{enumerate}
\item Sort through the  Monitored Source List, picking out sources that flared prior to 2017 and had significant and visually interesting activity, such as one or more well-defined flares. We found that 110 out of 158 were of interest.  We excluded later flares, because publications for those flares were unlikely to have been complete by mid-2018. 
\item Determine if the {\it Fermi} sources had {\it Swift} ToO follow-ups by searching through the online {\it Swift} ToO archive at  {\it www.swift.psu.edu/secure/toop/summary.php}. It was found that 91 out of 110 {\it Fermi}-LAT sources have a least one {\it Swift} observation after {\it Fermi}'s 2008 launch. 
\item Search for publications using {\it Fermi} and {\it Swift} data for individual sources in this collection, using the SAO/NASA ADS system: {\it adsabs.harvard.edu}.  Of the 91 sources with observations by both satellites, 32 had publications using data from both, and 21 of these explicitly mentioned the {\it Swift} ToO.  A list of these references is given in  Appendix A. 
\item  For all 110 sources of interest, download the FITS files of the gamma-ray data, then use python code to overplot light curves for similar types of sources (e.g. BL Lac objects) to look for similarities.  An example is shown in Fig. \ref{fig:BLLacs}.
\item Derive a quantitative representation of the statistically significant structures (height and duration) in the light curves, using a Bayesian Block analysis \citep{Scargle}.  The resulting light curve for PKS 0426$-$380 is shown in Fig. \ref{fig:Bayesian}.
\item Categorize the sources: type of object, peak daily flux, duration of flares, and number of flares. 
\end{enumerate}

\smallskip
\noindent {\bf 3. Results}
\vspace{+0.025in}

For the characteristics of the ``successful'' ToOs, we concentrate on the 21 that explicitly mentioned the {\it Swift} ToO in the publication.  One was a flare of the Crab Nebula.  We exclude that as being a special case.  Some results:

\begin{itemize}
\item Approximately 2/3 of the successful ToOs involved blazars that were historically active, showing multiple flares over an extended time range.  The rest had only one or a few flares. 
\item Over half of these results involved Flat Spectrum Radio Quasars (FSRQs), with the rest divided between BL Lac objects and blazars of uncertain type. This result is not surprising, since gamma-ray FSRQs are typically more variable than BL Lac objects \citep{3LAC}.
\item Flare durations were widely scattered, ranging from 1 day to 9 months, with an average duration of $\sim$38 days. 
\end{itemize}

Comparing the sources that produced publications with those that did not gives the following:

\begin{itemize}
\item Of the 8 {\it Fermi}-LAT sources with peak flare $>$ 8 $\times$ $10^{-6} \rm{ph}~\rm{cm}^{-2}~s^{-1}$ above 100 MeV, 7 resulted in publications including {\it Swift} ToO results.  The one remaining case was a flare of 3C273 that occurred at a time {\it Swift} had a sun angle constraint and was not able to execute the ToO.  
\item For {\it Fermi}-LAT sources with peak flare 1--3  $\times$ $10^{-6} \rm{ph}~\rm{cm}^{-2}~s^{-1}$ above 100 MeV, 14 out of 76 cases had publications including {\it Swift} ToO results.
\item Sources that resulted in publications had an average of 2.7 flares in the {\it Fermi}-LAT data, while those that were not published had an average of 1.7 flares. 
\end{itemize}

The peak flux for these flaring sources is clearly an important parameter.  Another way to express the flux dependence is:  for {\it Fermi}/{\it Swift} published sources, the median flux was 2.4 $\times$ $10^{-6} \rm{ph}~\rm{cm}^{-2}~s^{-1}$ above 100 MeV; for {\it Fermi}/{\it Swift} unpublished sources, the median flux was 2.0 $\times$ $10^{-6}$ in the same units; and for sources that did not have a {\it Swift} ToO, the median flux was 1.1 $\times$ $10^{-6}$ in the same units.

\smallskip
\noindent {\bf 4. Conclusions}
\vspace{+0.025in}
%
%

%
Based on a study of {\it Swift} ToOs for flaring {\it Fermi}-LAT gamma-ray sources, we can draw these conclusions:

\begin{itemize}
\item Peak gamma-ray flux is the clearest indicator of which {\it Swift} ToOs are most likely to result in a publication.  Very bright flares are almost always published. 
\item Historically active blazars, ones with multiple flares over the time of the {\it Fermi} mission, are more likely to result in publications than those with only a few flares. 
\item The type of source and the durations of flares have little effect on whether a source produces a publication.
\end{itemize}

While these conclusions offer guidance about {\it Swift} ToO requests, they are not the only considerations.  Higher flux flares are rarer and arguably more interesting than lower flux flares. However, lower flux flares should not be ruled out every time because there are  instances where lower flux flares provide groundbreaking insight, such as  the indication of neutrino emission from TXS 0506+056 \citep{0506}.  Since we have more information on the historic sources, this obviously will lead to more publications. More data on a source means more to analyze, which leads to more conclusions on the source mechanisms. Again the TXS 0506+056 offers a counterexample, since it is not a particularly active gamma-ray source. 

One other result from this study stands out.  Our searches turned up no obvious publications using data from both satellites for a large fraction of the flaring {\it Fermi}-LAT gamma-ray sources that also have {\it Swift} data.  We may, of course, have missed some references, or they could still be in preparation.  Some of the sources also have publications that do not use both data sets.  Nevertheless there appears to be a significant body of observational data about flaring blazars that has not yet been exploited.  Information about these sources is found in Appendix B.

\smallskip
\noindent {\bf 5. Acknowledements}
\vspace{+0.025in}
%
We wholeheartedly acknowledge and thank the hardworking members of the {\it Fermi}-LAT Collaboration and the {\it Swift} team for providing the data for this study. We thank and recognize postdoctoral NASA fellow Sara Buson for her help in creating Bayesian Block python code used in the study for flare amount, duration, and variation analysis. We also recognize Penn State University and their documentation of {\it Swift} ToO requests, Strasbourg University for their SIMBAD astronomical database used to help classify sources, as well as the Smithsonian Astrophysical Observatory and NASA efforts in providing the Astrophysics Data System used in searching for targeted publications.

{}

\vspace{+0.025in}
\vspace{-0.6cm}

\normalsize

\smallskip
\noindent {\bf Appendix  A}
\vspace{+0.025in}

Publications resulting from {\it Fermi}-LAT/{\it Swift} ToO studies are given below. 

Abdo A. A., et al. 2015, ArXiv:1411.4915v2

Abdo, A. A., et al. 2010, ArXiv:1011.1053v1

Abdo, A. A., et al. 2010, ArXiv:0912.4029v2

Abdo, A. A., et al. 2009, ArXiv:0905.4558v1

Abdo, A. A., et al. 2009, ArXiv:0903.1713v1

Abramowski, A., et al. 2016, ArXiv:1409.0253v1

Ackermann, M., et al. 2014, The Astrophysical Journal, Volume 	786(157), 17

Aleksic, J., et al. 2018, ArXiv:1401.5646v2

Bottacini, E., Bottecher, M., Pian, E., Collmar, W., 2016, 	arXiv:1610.01617v1

D?Ammando, F., et al. 2015, ArXiv:1504.05595v1

D?Ammando, F., et al. 2014, ArXiv:1410.7144v1

D?Ammando, F., et al. 2013a, ArXiv:1302.5439v1

D?Ammando, F., et al. 2012, ArXiv:1209.0479v1

D?Ammando, F., Orienti, M., 2015, ArXiv:1510.06416v1

Dutka, M., et al. 2016, ArXiv:1612.08061v1

Ghisellini G., Tavecchio, F., Foschini, L., Bonnoli, G.,Tagliferri, G., 2013, 	ArXiv:1302.4444v2

Kaur, N., Baliyan, S. K., 2018, ArXiv:1805.04692v1

Larionov, V. M., et al. 2016, ArXiv:1606.07836v1

Pian, E., et al. 2018, ArXiv:1011.3224v1

Piano, G., et al. 2018, ArXiv:1805.05640v2

Pittori, C., et al. 2018, ArXiv:1803.07529v1

Pucella, G., et al. 2010, A\&A 522, A109

Rani, B., et al. 2016, ArXiv:1609.04024v1

Tagliaferri, G., et al. 2015, arXiv:1503.04848v2

Tanaka, Y. T., et al. 2016, ArXiv:1604.05427v1

Weidinger, M., Ruger, M., Spanier, F., 2010, Astrophys. Space Sci. 	Trans., 6, 1?7

Zhang, J., et al. 2017, ArXiv:1709.02161v1

Zhang. S, Collmar. W, Torres. D. F, Wang. J, Lang. M, Zhang. S. N. 2010, 	A\&A, 514, A69

\noindent {\bf Appendix  B}
\vspace{+0.025in}

The tables list flaring sources from the {\it Fermi}-LAT Monitored Source List that had {\it Swift} observations but no obvious publication using the data from both satellites.  The flux values come from the automated analysis and are therefore approximate. 

\begin{table}
\caption{Sources with {\it Fermi}/{\it Swift} observations without a publication - Part 1 }
\begin{tabular}{l r r } \hline
\hline 
Name of Source  & Maximum daily flux & Number of flares       \\
\hline 
4C+01.02 / PKS 0106+01 & 3 & 3 \\
CGRaBS J0211+1051 & 1.1 & 1   \\
S3 0218+35 & 4 & 4   \\
4C+28.07 & 2.2 & 3   \\
PKS 0301$-$243 & 1.4 & 1   \\
PKS 0336$-$01 & 2 & 2  \\
PKS 0402$-$362 & 6 & 4   \\
NRAO 190& 3 & 3   \\
PKS 0454$-$234 & 1.7 & 4   \\
PKS 0458$-$02 & 1.9 & 2   \\
PKS 0502+049 & 3.2 & 2   \\
PKS 0507+17 & 4.1 & 2   \\
VER 0521+211 & 1 & 1   \\
PKS 0528+134 & 1.2 & 1   \\
OG 050 & 1.6 & 1   \\
B2 0619+33 & 1.8 & 1   \\
4C 14.23 & 2.1 & 1   \\
PKS 0727$-$11 & 1 & 1   \\
PKS 0736+01 & 2.1 & 2   \\
PKS 0805$-$07 & 1.6 & 2  \\
0827+243 & 1.2 & 1   \\
PKS B0906+015 & 1.5 & 2   \\
S4 1030+61 & 1.3 & 1   \\
S5 1044+71 & 1.3 & 2   \\
1150+497 & 2.4 & 1   \\
Ton 599 & 2.1 & 3   \\
ON 246 & 1 & 1   \\
GB6 B1310+4844 & 2.1 & 1   \\
PKS 1313$-$333 & 1.3 & 1   \\
PKS 1329$-$049 & 4.1 & 1   \\
B3 1343+451 & 1.3 & 1   \\
PKS 1424-41 & 3 & 5   \\
B2 1520+31 & 2 & 2   \\
TXS 1530$-$131 & 1.5 & 1   \\
PKS 1622$-$253 & 3 & 2   \\
GB6 J1700+6830 & 1 & 1   \\
1730$-$130 & 1.5 & 1   \\
OT 081 & 4 & 1   \\
S5 1803+78 & 1.3 & 1  \\
S4 1800+44 & 3 & 1   \\
  \\\hline 
\end{tabular}\\
Note: the flux values are approximate and in units of  $10^{-6} \rm{ph}~\rm{cm}^{-2}~s^{-1}$ above 100 MeV.\\
\label{tab1}
\end{table} 

\begin{table}
\caption{Sources with {\it Fermi}/{\it Swift} observations without a publication - Part 2 }
\begin{tabular}{l r r } \hline
\hline 
Name of Source  & Maximum daily flux & Number of flares       \\
\hline 
PKS 1824$-$582 & 3.9 & 2   \\
CGRaBS J1848+3219 & 2 & 2   \\
B2 1846+32B & 2.2 & 1   \\
PKS 2032+107 & 3.2 & 1   \\
PKS 2136$-$642 & 1.5 & 1   \\
NRAO 676 & 4.5 & 3   \\
BL Lac & 2.3 & 4   \\
PKS 2233$-$148 & 3 & 2   \\
TXS 2241+406 & 1.5 & 1   \\
B2 2308+34 & 1.5 & 2   \\
PKS 2320$-$035 & 1 & 1   \\
  \\\hline 
\end{tabular}\\
Note: the flux values are approximate and in units of  $10^{-6} \rm{ph}~\rm{cm}^{-2}~s^{-1}$ above 100 MeV.\\
\label{tab1}
\end{table} 

\end{document}